\documentclass[prc,preprint,superscriptaddress,nofootinbib,showpacs]{revtex4}
\usepackage{amsfonts}
\usepackage{amssymb}
\usepackage{amsfonts}
\usepackage{amsmath}
\usepackage{amssymb}
\usepackage{amsthm}
\usepackage{epsf}
\usepackage[dvips]{graphicx}

\begin{document}

\title{Analytical properties of the gluon propagator from truncated Dyson-Schwinger
equation in complex Euclidean space}

 \author{L.~P. Kaptari\footnote{On leave from  Bogoliubov Lab.~Theor.~Phys., 141980, JINR, Dubna, Russia }}
\affiliation{Institute of Modern Physics, Chinese Academy of Science, 509 Nanchang Road, 730000, Lanzhou, China}
\author { B.~K\"ampfer}
\affiliation{Helmholtz-Zentrum Dresden-Rossendorf, PF 510119, 01314
Dresden, Germany}
\affiliation{Institut f\"ur Theoretische Physik, TU Dresden, 01062 Dresden, Germany}

\author { P. Zhang}
\affiliation{Institute of Modern Physics, Chinese Academy of Science, 509 Nanchang Road, 730000, Lanzhou, China}

\begin{abstract}
We suggest a framework based on the rainbow approximation
with effective parameters  adjusted to lattice data. The analytic
 structure of the gluon and ghost propagators of QCD in Landau gauge
is analyzed by means of  numerical solutions of the coupled system of truncated Dyson-Schwinger equations.
 We find that the gluon and ghost dressing functions
 are singular in complex Euclidean space with singularities as isolated
 pairwise conjugated poles. These poles  hamper solving numerically the Bethe-Salpeter equation
 for glueballs as bound states of two interacting dressed gluons.
  Nevertheless, we argue that, by knowing the position of the poles and their
 residues,  a reliable algorithm for numerical  solving the Bethe-Salpeter equation
  can be established.

\end{abstract}
 
\maketitle
\section{Introduction}

 Due to the  non-Abelian and confinement properties of
 Quantum Chromodynamics (QCD), gluons  obeying  self-interactions
 can form colorless  pure gluonic bound states, also referred to as glueballs.
The occurence of glueballs is one of the early predictions of the  strong
interactions described by  QCD~\cite{Jaffe}.
 However, despite many years of
experimental efforts, none of these gluonic states have been established unambiguously,
cf. Ref.~\cite{Jia:2016cgl}. Possible
reasons for this include the mixing between glueballs and "conventional" mesons, the lack of solid
information on the glueball production mechanism, and the lack of knowledge about glueball
decay properties. Therefore,   study of glueballs is one the most interesting and
challenging problems intensively studied by theorists and experimentalists;
 a bulk of the running and projected   experiments   of the  research centers,
 Belle (Japan),  BESIII (Beijing, China),
 LHC (CERN), GlueX (JLAB,USA), NICA (Dubna, Russia), HIAF (China),  FAIR (GSI, Germany) etc.,
include  in the research programms
comprehensive investigations of possible manifestations of glueballs.
Theoretically,  there are several approaches in
studying  glueballs. One can mention phenomenological models mimicking certain nonperturbative
QCD aspects, such  as the flux tube model~\cite{Robson:1978iu,Isgur:1984bm},
constituent models~\cite{Jaffe,Carlson:1984wq,Chanowitz:1982qj,Cornwall:1982zn,Cho:2015rsa,Boulanger:2008aj},
holographic approaches~\cite{Bellantuono:2015fia,Chen:2015zhh,Brunner:2016ygk},
approaches based on QCD Sum Rules~\cite{Shifman:1978bx,Shuryak:1982dp,kolya,kolya1,KolyaPRL}.
"Experimental" studies are   performed within the Lattice QCD (LQCD)
approaches~\cite{Albanese,Chen,Morningstar,Gabadadze}
(for a more detailed  review see Ref.~\cite{Glueballstatus} and references therein).
It should be noted that these theoretical
approaches provide  values of glueball masses which can
differ from each other as much as 1 GeV and even more.
No single approach has consistently reproduced lattice gauge calculations,
 cf. Refs.~\cite{Albanese,Morningstar,Chen,Gabadadze}.
One can assert only that  the consensus of the past
two decades from lattice gauge theory  and theoretical predictions
 is that the lightest glueball is a scalar
($J^{PC} = 0^{++}$) state in the 1.5-1.8 GeV mass range, accompanied
by a tensor ($J^{PC} =2^{++}$) state above 2~GeV.

 Another interesting problem is the glueball-meson
mixing in the lowest-lying scalar mesons.  The question whether the lowest-lying scalar mesons
are of a pure quarconium nature, or  there are   mixing  phenomena of glueball  states~\cite{mixing} remains still open.
To solve these problems one needs to develop models  within which it becomes possible to investigate, on a
common footing, the glueball masses, glueball wave functions, decay modes and constants,  etc.
Such approaches can be based on the combined Dyson-Schwinger (DS) and Bethe-Salpeter (BS)
formalisms, cf. Refs.~\cite{glueBS,GlueBallBSPRD87}. It is worth mentioning that theoretically  such models,
with direct calculations of the corresponding diagrams, encounter difficulties
in solving the DS equation, related to divergencies of
loop integrals and to the theoretical constrains on gluon-ghost and gluon-gluon vertices,
such as Slavnov-Taylor identities. These circumstances
  result in rather cumbersome expressions for the DS equation,
  hindering straightforward numerical calculations.

In the present paper we suggest an approach, similar to the rainbow  Dyson-Schwinger-Bethe-Salpeter
model for quark propagators~\cite{rob-1}, to solve the DS equation for gluon and ghost
propagator with effective rainbow kernels.
The formidable
success of the rainbow approximation for quarks  in  describing  mesons as quark-antiquark
 bound states within the framework of the BS equation
 with momentum dependent quark mass functions, determined directly by the DS  equation, such as meson masses
 \cite{rob-1,Maris:1999nt,Maris:2003vk,Holl:2004fr,Blank:2011ha},   electromagnetic properties
 of pseudoscalar mesons~\cite{Jarecke:2002xd,Krassnigg:2004if,Roberts:1994hh,Roberts:2007jh})
 and  other   observables~\cite{ourFB,wilson,rob-2,Alkofer,fisher},
  persuades us  that the rainbow-like approximation may be successfully applied
  to gluons, ghosts and glueballs as well.
 The key property of such a framework is the self-consistency of the  treatment of the
 quark and gluon propagators in both, DS  and BS equations by employing in both cases the same
 approximate interaction kernel.
 Recall that the rainbow model for quarks consists of a replacement of the product
 of the coupling  $g$ dressed gluon propagator ${\cal D}^{ab}_{\mu\nu}(k^2)$  and  dressed
 quark-gluon vertex $\Gamma_\nu$ by an effective running coupling and by the free vertex
 $\Gamma_\nu^0$~\cite{rob-1,rob-2},

 \begin{equation}
 \displaystyle\frac{g^2}{4\pi} {\cal D}_{\mu\nu}^{ab}(k^2)
 \Gamma_\nu \to \displaystyle\frac{{\cal Z}(k^2)}{k^2} D_{\mu\nu}^{free}\Gamma_\nu^{(0)},
\label{rain}
 \end{equation}
 where $a,b$ are color indices and  ${\cal Z}(k^2)$ is the  effective  rainbow  running coupling.
The explicit form of ${\cal Z}(k^2)$  has been induced by the fact that,
in the Landau gauge, it is proportional to the nonperturbative running coupling $\alpha_s(k^2)$
which, in turn,  is determined by  the gluon $Z(k^2)$ and ghost $G(k^2)$ dressing
functions~\cite{AlkoferRunning,Alkofer_FischerRunning,FischerPhD,physRep,IRGluonProp_AlkoferSmekal,AlkoferEstrada,RunningCouplingAkinston,alphaSNonpert,IRGreen_FewBody2012} as
\begin{equation}
\alpha_s(k^2) = \frac{g^2}{4\pi} G^2(k^2,\mu^2)Z(k^2,\mu^2),
\label{alphas}
\end{equation}
where  $\mu^2$ is a renormalization scale parameter at $k^2=\mu^2$ whith $ G^2(k^2,k^2)Z(k^2,k^2)=1$.
In what follows, the parameter  $\mu^2$ is suppressed in our notation and a simple notation  $G(k^2)$ and $Z(k^2)$ is used
for the dressing functions.

   In principle, if one were able to solve the DS equation,
 the approach would  not depend on  any additional parameters. However, due to  known technical problems,
 one restricts oneself to  calculations of the few first terms of the perturbative series,
 usually within  one-loop approximation, thus arriving at  the   truncated Dyson-Schwinger (tDS)
 and truncated Bethe-Salpeter (tBS) equations,
  known as the rainbow-ladder approximation.
 The merit of such an approach is that, once the effective parameters are fixed,
 the whole spectrum of the tBS bound states is supposed to  be described without additional
approximations.

In the present paper we investigate  the prerequisites to the interaction
kernel of the combined Dyson-Schwinger and Bethe-Salpeter  formalisms to be used
in subsequent calculations to describe the  glueball mass spectrum.
  Note that  within such an approach it becomes
possible to theoretically investigate  not only the mass spectrum of glueballs, but also
different processes of their decay, which are directly connected with fundamental QCD problems
(e.g., $U(1)$ axial anomaly, transition form factors etc.) and with the challenging  problem of
 changes of hadron matter  characteristics at finite temperatures and densities.   All these circumstances  require  an
adequate theoretical foundation to describe the glueball mass spectrum and their  covariant wave functions
 (i.e. the tBS partial amplitudes) needed in   calculations
of the relevant  Feynman diagrams and observables.

   Due to   the
  momentum dependence of the gluon and ghost dressing functions,  the tBS equation requires
  an analytical continuation of the gluon and ghost propagators in the
  complex plane of Euclidean momenta which can be achieved either by corresponding numerical
  continuations of the solution obtained along the positive real axis or by solving directly
  the DS equation in the complex domain of validity
  of the equation itself. An analysis of the analytical properties of the propagators
  is of crucial importance, since if they are   singular functions,
  the numerical calculations of corresponding integrals can be essentially hampered or even
  impossible in such a case. We perform a detailed analysis of  the tDS equations solution
   by  combining the   Rouch\'e's and Cauchy's theorems. Since the main goal of our analysis
   is the subsequent  use of the
   gluon and ghost  propagator  functions evaluated at such complex
   momenta for which they are needed in the tBS equation, we
   focus our attention on this region of Euclidean space.

 Our paper is organized as follows. In Sec.~\ref{s:bse}, Subsecs.~\ref{Bet}~and~\ref{Dys},
 we briefly discuss the  tBS and tDS equations, relevant to describe a glueball as two-gluon
 bound states. The rainbow approximation for the tDS equation kernel is  introduced
 in Subsec.~\ref{bow}, and  the numerical solutions of the tDS together with
 comparison with lattice QCD data are presented in Subsec.~\ref{real}.
  Section~\ref{sec3}
 is entirely dedicated to the solution of the   tDS equation for complex
 Euclidean momenta, where the solutions are sought.
 The analytical structure  of the gluon and ghost propagators in complex Euclidean space
is discussed in Subsec.~\ref{analyt}.
 It is found that   the ghost dressing function is analytical
  in the right hemisphere and contains pole-like singularities in the left
 hemisphere  of complex momenta squared, $k^2$, while the gluon dressing function contains singularities
 in the entire Euclidean space.
  A thorough investigation of the pole structure of propagators is   presented in
 Subsec.~\ref{poleStruc}. By combining    Rouch\'e's and Cauchy's theorems, the position of first
 few poles and the corresponding residues  of the
dressing functions  are found with a good accuracy. This information is
useful in elaborating adequate algorithms for numerically solving the tBS equations in presence of
pole-like singularities.  Conclusions and summary are collected
in Sec.~\ref{summary}. In the Appendix~\ref{app}, the
behaviour of the nonperturbative running coupling is discussed in connection with the choice of the
rainbow kernels and a parametrization of the lattice QCD data in form  of a sum of Gaussian terms is presented.

\section{Bethe-Salpeter and Dyson-Schwinger Equations}
\label{s:bse}

As mentioned above,  our ultimate goal is to elaborate an effective model,
based on the Dyson-Schwinger-Bethe-Salpeter approach, to
 describe a glueball made from two gluons that
are solutions of the tDS equation for the gluon propagator. As in the
rainbow approach~\cite{Alkofer,rob-1,rob-2}, a central requirement of our model
is the self-consistent  treatment of the gluon propagator in both, tBS  and tDS equations.
In the following we work along this strategy,
i.e. we  elaborate an effective model
 within which  (i) the solution of the gluon and ghost propagators, consistent with
  the lattice  data, is sought  along the positive real axis
 of the momentum, (ii) then  the real solution is generalized
 for  complex momenta, relevant to the domain in
 Euclidean space where the tBS is defined, and (iii) an analysis, regarding the analytical properties
 of the complex solution, can be performed.

 \subsection{Bethe-Salpeter equation}
\label{Bet}
\begin{figure}[!ht]
 \includegraphics[scale=0.4 ,angle=0]{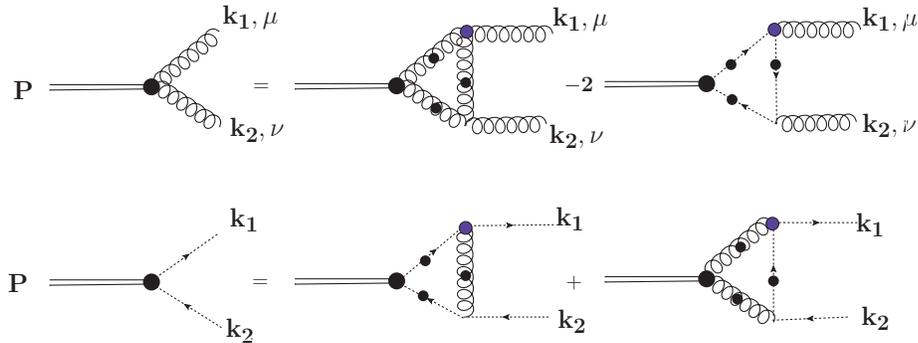}\vspace*{3mm}
 \caption{ Diagrammatic representation  of the Bethe-Salpeter  equations for gluon (wiggly lines)
  and ghost (dashed lines) bound states.  The irreducible one-particle vertices and the full propagators are represented by filled blobs.}
 \label{Bseq}
 \end{figure}

 We are working in Landau gauge and, consequently, we need
to take into account the contribution of the Faddeev-Popov ghosts. Thus, one
needs a generalization of the usual BS  scheme that allows
for mixing of bound states of different fields.  In general, the completye system of
BS equations includes also the contribution of quark-antiquark bound states, i.e.
involves also glueball-meson mixing in the BS calculations. The problem of how large can be
these mixing effects is not yet clearly settled. However, there are some indications, based on lattice
calculations of the   pure glue pseudoscalar glueball~\cite{glueSpectrum}, that at least in the pseudoscalar channel the
glueball-meson mixing can be neglected, see also the discussion in Ref.~\cite{glueBS}.
 In what follows we will be interested in bound states for a pure gauge theory, that is neglecting quarks. The corresponding
 system of coupled tBS equations is presented diagrammatically in Fig.~\ref{Bseq}.  The explicit
  form  of the corresponding equations can be found, e.g.  in Ref.~\cite{GlueBallBSPRD87}.

 To solve this system numerically, we need reliable information on the nonperturbative propagators
of ghosts and gluons and their analytical properties  in  complex Euclidean momentum space.
  This can be achieved, e.g.  by solving the tDS  coupled equations for the gluon and ghost propagators along
  the real axis of the  momentum $k$ and then to use the tDS equation at
 complex external momenta $k$ in Euclidean space.
\subsection{Dyson-Schwinger equation}

\label{Dys}
 \begin{figure}[!ht]
 \includegraphics[scale=0.55 ,angle=0]{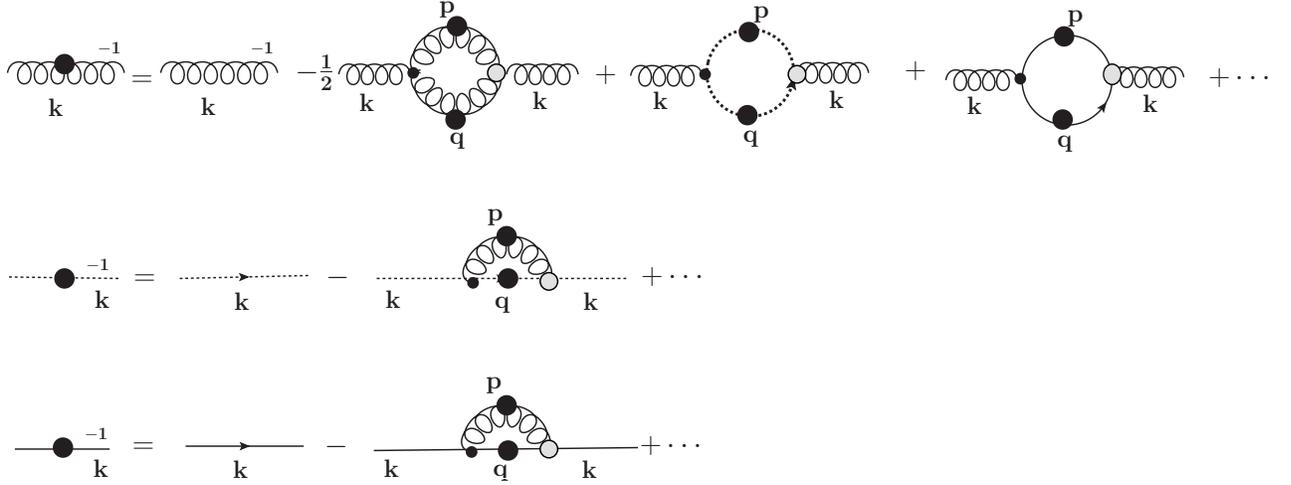}\vspace*{3mm}
 \caption{ Diagrammatic representation  of the  Dyson-Schwinger equations for gluon,
  ghost and quark propagator. The internal wiggly,  dashed and solid lines denote
   the full propagators represented by filled blobs. The irreducible one-particle
   vertices are denoted by open circles. In the gluon Dyson-Schwinger equation terms with four-gluon vertices
  have been dismissed.}
 \label{dse}
 \end{figure}

The coupled equations of the quark, ghost and gluon propagators,  and the corresponding vertex
functions  are often considered as integral formulation being equivalent to full QCD. While
there are attempts to solve this coupled set of DS equations by sutable numerical procedures, for certain purposes
some approximations and  truncations~\cite{fisher,Maris:2003vk,physRep} of the
exact interactions  are appropriate.  This leads to the  truncated Dyson-Schwinger  system of
coupled equations for the gluon, ghost and quark propagators as depicted in Fig.~\ref{dse}.
As mentioned above, accounting for the quark loop diagrams in the full  tDS equation results in a gluon-quark mixing
in the tDS equation and in a glueball-meson mixing in tBS. In most calculations
  such   a mixing  is neglected in the tBS  equations. For the sake of consistency and in order
 to reduce the number of phenomenological parameters of the approach, the  quark loops
in our approach are neglected as well. This can be justified by the
 observation~\cite{fisher,LatticeQuenchedvsUnquenched,QuenchedUnquenchedGluon} that
in the tDS equation the unquenched effects are rather small in
 the dynamical quark masses. As for  the gluon propagator, such effects  are seen only
 in the neighbourhood of the gluon bump, $k\sim 0.85-1.0 \, {\rm  GeV/c}$, where
  the  screening effect from the  creation of quark-antiquark pairs from the vacuum
  slightly decreases  the value of the gluon dressing around its maximum. In our approach this
  effect is implicitly taken into account by adjusting the phenomenological parameters of the model
  to the full, unquenched  lattice calculations~\cite{GhostLatticeMishaPRD,BornyakovLattice}.

  In the Landau gauge  the gluon propagator ${\cal D}_{\mu\nu}^{ab}(k)$ and
  ghost propagator ${\cal D}_G^{ab}(k)$ are expressed via the dressing functions $Z(k)$ and $G(k)$  as
 \begin{eqnarray}&&
 {\cal D}_{\mu\nu}^{ab}(k) = -i\delta^{ab}{ D}_{\mu\nu}=- i\delta^{ab}\frac{Z(k)}{k^2}t_{\mu\nu}(k), \\ &&
 {\cal D}_G^{ab}(k)=i\delta^{ab} D_G(k)=i\delta^{ab}\frac{G(k)}{k^2},
 \end{eqnarray}
 where $t_{\mu\nu}(k)$ is the transverse projection operator,
 $t_{\mu\nu}(k)=g_{\mu\nu}-\displaystyle\frac{k_\mu k_\nu}{k^2}$.
 Then the  corresponding dressing functions obey  the   tDS equation (cf. Fig.~\ref{dse}).
\begin{eqnarray}
G^{-1}(k^2)&=& \tilde Z_3  +  \frac {3}{ 4\pi^3 k^2}i \int
d^4 q   \left[\frac{g^2}{4\pi} \Gamma_{\mu}^{(0)}(q) D^{\mu\nu}(p^2) \Gamma_{\nu}(k,q,p) \right]
 \frac{G(q^2)}{q^2},
\label{sdeGhost}\\[3mm]
Z^{-1}(k^2)&=&Z_3+\frac {i}{ 8\pi ^3 k^2}\int d^4 q  \frac{Z(q^2)}{q^2}
\left[\frac{g^2}{4\pi} \Gamma_{\mu\rho\alpha}^{(0)}(k,p,-q)
D^{\rho\sigma}(p^2)\Gamma_{\beta\sigma\nu}(q,-p,-k)\right]
t^{\alpha\beta}(q) t^{\mu\nu}(k)-\nonumber\\[3mm]
&-&\frac{i}{4\pi^3 k^2}\int d^4q \frac{G(q)}{q^2}
\Gamma^{(0)}_\mu(q)\left [\frac{g^2}{4\pi}D_G(p^2)\Gamma_\nu(p)\right]t^{\mu\nu}(k),
\label{sdeGluon}
\end{eqnarray}
where $p=q-k$ and $Z_3$ and $\tilde Z_3$ are the gluon and ghost renormalization  constants, respectively.
To solve this system of equations one needs information about the
three-gluon vertex $\Gamma_{\beta\sigma\nu}$, the
gluon-ghost vertex $\Gamma_\nu$, the coupling $g$  and   the propagators
 ${\cal D}_{\mu\nu}^{ab}$ and ${\cal D}_G^{ab}$. The simplest approach
consists in a  replacement of the full dressed three-gluon and  ghost-gluon vertices by their bare values,
known as the Mandelstam approximation~\cite{CPCMandelstam,MANDELSTAM_Approx,PennigtonMandels}
and the y-max approximation~\cite{YmaxApprox}. In order to simplify the
angular integration, in the Mandelstam approximation the gluon-ghost coupling is  neglected.
 Then the resulting solution exhibits a  rather singular gluon propagator at the origin.
 In Ref.~\cite{YmaxApprox} the coupling of the gluon to the ghost was not neglected,
 however  additional simplifications for $Z(k^2)$ and $G(k^2)$ have been
 introduced, again  to  facilitate the angular integrations  and the analytical and numerical
analysis of the equations. From these calculations it has been concluded that
it is not the gluon, but rather the ghost propagator that is highly singular in the
deep infrared limit. A more rigorous analysis of the tDS equation has been presented  in a series of
 publictions (see, e.g. Refs.~\cite{IRGluonProp_AlkoferSmekal,SmekalAnnPhys,Fischer_Alkofer_Reinhard,FischerPhD}
 and references therein), where much attention has been focused on a detailed investigation
 of the gluon-gluon and ghost-gluon vertices and on the  implementation of
 the Slavnov-Taylor identities for these  vertices.
 With some additional approximations the infrared behavior of gluon and ghost propagators
has been obtained analytically and compared with the available lattice calculations.
In Ref.~\cite{PawlowskyFicher} a thorough analysis of the relevance of the Slavnov-Taylor identities,
 renormalization procedures and divergences in the tDS equation is presented in some details.
 Comparison of the numerical calculations
for the gluon and ghost dressing functions and  running coupling $\alpha_s$ with lattice data have
been presented as well. Similar calculations together with a comparison with  lattice data are
presented also in Ref.~\cite{IRGreen_FewBody2012}
(for a more detailed review see Ref.~\cite{FisherReview} and references therein quoted).
It should be noted that   the above quted approaches result in rather cumbersome expressions for
the system of tDS equations which, consequently, cause  difficulties in finding the numerical solutions.
Yet, a direct generalization to complex Euclidean space becomes problematic
due to numerical problems at large $|k^2|$ of the complex  momentum.
\subsection{Rainbow approximation for ghosts and gluons}
\label{bow}
In the present paper we suggest an approximation
for the interaction kernels in Eqs.~(\ref{sdeGhost}) and (\ref{sdeGluon}),
similar to the rainbow model~\cite{Alkofer,rob-1,rob-2}, Eq.~(\ref{rain}),   which
allows for an analytical angular integration  in the gluon and ghost loops and facilitates
further numerical calculations for the complex momenta.  The results of the
lattice calculations of the running coupling $\alpha_s(k^2)$,  Eq.~(\ref{alphas}),
can serve as a guideline in choosing the explicit form of these kernels. The gist of our approximations is as follows:
\begin{eqnarray}&&  \!\!\!\!\!\!\!\!\!\!\!\!\!\!
\left[\frac{g^2}{4\pi} \Gamma_{\mu}^{(0)}(q) D^{\mu\nu}(p^2) \Gamma_{\nu}(k,q,p) \right]
= \Gamma_{\mu}^{(0)}(q)t^{\mu\nu}(p)\Gamma_{\nu}^{(0)}(k) F^{eff}_1(p^2),
\label{rain1}\\[3mm] && \!\!\!\!\!\!\!\!\!\!\!\!\!\!
\left[\frac{g^2}{4\pi} \Gamma_{\mu\rho\alpha}^{(0)}(k,p,-q)
D^{\rho\sigma}(p^2)\Gamma_{\beta\sigma\nu}(q,-p,-k)\right] =A
  \Gamma_{\mu\rho\alpha}^{(0)}(k,p,-q) t^{\rho\sigma}(p)\Gamma_{\beta\sigma\nu}^{(0)}(q,-p,-k)F^{eff}_1(p^2),
  \label{rain2}\\[3mm] &&
\!\!\!\!\!\!\!\!\!\!\!\!\!\!
 \left [\frac{g^2}{4\pi}D_G(p^2)\Gamma_\nu(p)\right]=\Gamma_\nu^{(0)}(p)F^{eff}_2(p^2), \label{rain3}
\end{eqnarray}
where $A\sim 1/3$ is a phenomenological parameter which takes into account the difference in
normalizations of the gluon and ghost vertices.
The effective form-factors $F^{eff}_{1,2}(p^2)$ are proportional to the
infra-red part of the  running coupling   $\alpha_s(k^2)$, see~Eq.~(\ref{alphas}).
The   free gluon-gluon vertices $\Gamma_{\mu\rho\alpha}^{(0)}$ and $\Gamma_{\beta\sigma\nu}^{(0)}$ as well as
 the ghost-gluon vertices  $\Gamma^{(0)}_\mu(q)$ and $\Gamma^{(0)}_\nu(q)$ read
in the Landau gauge
\begin{eqnarray}  &&
\Gamma_{\mu\rho\alpha}^{(0)}(k,p,-q) = 2k_\alpha g_{\mu\rho} +2q_\mu g_{\rho\alpha} -2k_\rho g_{\alpha\mu},
\label{vertexA} \\ &&
\Gamma_{\beta\sigma\nu}^{(0)}(q,-p,-k)=2q_\nu g_{\beta\sigma} + 2k_\beta g_{\nu\sigma}-2k_\sigma g_{\beta\nu},
\label{vertexB} \\ &&
\Gamma^{(0)}_\mu(q) = -q_\mu; \qquad \Gamma^{(0)}_\nu(p)=-p_\nu =-(q-k)_\nu =-q_\nu\label{vertexC}.
\end{eqnarray}

 The rainbow approximation for the propagators in Minkowski space is obtained by
inserting Eqs.~(\ref{rain1})-(\ref{vertexB}) in to Eqs.~(\ref{sdeGhost}) and (\ref{sdeGluon})
and by contracting  the Lorenz indices.
Further calculations are performed in Euclidean space. For this, we
perform the Wick rotation of the loop integrals and specify   explicitly the form of
$F^{eff}_{1,2}(p^2)$.  Since we envisage the  further use of
the tDS equation solution in the tBS equation for  bound
states, where the main contribution comes from the IR region,
the perturbative ultra-violet (UV) parts of $F_{1,2}^{eff}$ are neglected. Such an approximation
corresponds rather to the AWW kernel~\cite{Alkofer} than to the full Maris-Tandy model~\cite{rob-1}.
As in the case of the quark rainbow
approximation~\cite{rob-1,rob-2,tandy1,wilson,physRep}, the explicit form of $F^{eff}_{1,2}(p^2)$
is inspired by the fact that the r.h.s. of Eqs.~(\ref{rain1})-(\ref{rain3}) are proportional to the
running coupling~(\ref{alphas}). The available  QCD lattice results~\cite{GhostLatticeMishaPRD}
show that, in the deep IR region, $\alpha_s$ increases as $k^2$ increases and reaches its maximum value at
$k\sim 0.8-0.9\  {\rm GeV/c}$; then it decreases as $k^2$ increases  and acquires
the perturbative behaviour in the UV region. In
Ref.~\cite{GhostLatticeMishaPRD} an interpolation formula consisting of three terms (monopole, dipole and
quadrupole, multiplied by $k^2$) has been proposed to fit the data. However, we prefer  an interpolation
formula which, in our subsequent   calculations, allows  to perform angular integrations
 analytically and assures a good convergence of the loop integrals.  For
 this  we use a Gaussian interpolation formula and
refitted  the lattice data~\cite{GhostLatticeMishaPRD} in the IR region
with several Gaussian terms and achieved a good agreement with data (see Appendix).  This stimulate
us to use for  $F^{eff}_{1,2}(p^2)$  the same interpolation formulae.
We found that  one Gaussian term for $F^{eff}_{2}(p^2)$ and two terms for $F^{eff}_{1}(p^2)$
 are quite sufficient
 to obtain a reliable solution of Eqs.~(\ref{sdeGhost})-(\ref{sdeGluon}):
\begin{eqnarray} &&
F^{eff}_{1}(p^2)=  D_{1} \frac{p^2}{\omega_1^6} \exp{\left( -p^2/\omega_1^2\right)}+
D_{2} \frac{p^2}{\omega_2^6} \exp{\left( -p^2/\omega_2^2\right )} ,\label{ff1} \\ &&
F^{eff}_{2}(p^2)= D_{3} \frac{p^2}{\omega_3^6} \exp{\left(-p^2/\omega_3^2\right)}. \label{ff2}.
\end{eqnarray}
With such a choice of the effective interaction,  the angular integration
can be  carried out analytically leaving one with a system of
one-dimensional integral equations in Euclidean space,
\begin{eqnarray}
G^{-1}(k^2) = \tilde Z_3
&-&
\frac{9}{8\pi} \sum_{i=1}^2 \frac{D_i}{k^2\omega_i^2}
\int   G(q^2) I_2^{(s)}\left(\frac{2q k }{\omega_i^2}\right) {\rm e}^{ -\frac{(q-k)^2}{\omega_i^2}} dq^2,
\label{eucl1}\\[3mm]
Z^{-1}(k)=Z_3
&-&
\frac{3A}{8\pi k^4}\sum_{i=1}^2\int dq^2 \frac{D_i}{\omega_i^2 q^2}Z(q){\rm e}^{ -\frac{(q-k)^2}{\omega_i^2}}
\left\{
I_1^{(s)}\left(\frac{2kq}{\omega_i^2}\right)
\left[\phantom{\!\!\!\!\!\frac12}
-12 k^3q -12kq^3-5kr\omega_i^2\right]\right.\nonumber \\[3mm]
&+&\left.
I_2^{(s)}\left(\frac{2kq}{\omega_i^2}\right)\left[\phantom{\!\!\!\!\frac12}6k^4 +6q^4+10\omega_i^4+18k^2q^2+
24k^2\omega_i^2+24 q^2\omega_i^2
\right]\right\} \nonumber \\[3mm]
 &+& \frac{3D_3}{8\pi k^4}\int dq^2 G(q)
 {\rm e}^{ -\frac{(q-k)^2}{\omega_3^2}}I_2^{(s)}\left(\frac{2kq}{\omega_3^2}\right),
 \label{eucl2}
\end{eqnarray}
where $I_n^{(s)}(x)$, with   $ x\equiv\displaystyle\frac{2kq}{\omega^2} $, are the  scaled
 (as emphasized by the label "(s)") modified Bessel
 functions  of the first kind defined as
 $I_{n}^{(s)}(x)\equiv \exp{(-x)}\ I_{n}(x)$.

\subsection{Numerical solution along the real axis}\label{real}

The resulting  system  of one-dimensional integral equations~(\ref{eucl1}) and  (\ref{eucl2})
we solve  numerically  by an iteration procedure. For this we discretize the loop integrals by
using the Gaussian integration formula, so that the system of integral equations reduces to a system
of algebraic equations. Independent parameters are
$\omega_i$ and  $D_i$, $i=1\cdots 3$, see Eqs.~(\ref{ff1}), (\ref{ff2}). We find that the iteration
procedure   converges  rather fast and  practically does not depend  on
the choice of the trial start functions. The phenomenological parameters
$\omega_i$ and  $D_i$ have been adjusted in such a way as to reproduce
 as close as possible the lattice QCD results~\cite{GhostLatticeMishaPRD,BornyakovLattice}.

\begin{figure}[hbt]             %
\includegraphics[scale=0.6,angle=0]{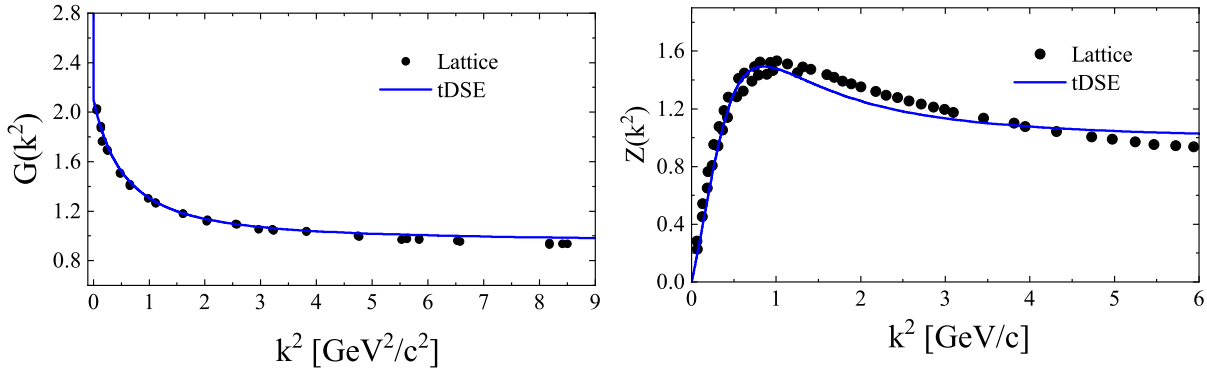}
\caption{(Color online) Solution of the  tDS equations~(\ref{eucl1})-(\ref{eucl2}) (solid lines) in comparison
with lattice calculations~\cite{BornyakovLattice,GhostLatticeMishaPRD}  (filled circles). Left panel:
ghost dressing function, right panel:  gluon dressing function.}
\label{ghost}
\end{figure}

Few remarks are in order here. First, the deep infrared behaviour of the
ghost and gluon propagators requires a separate consideration. It has been established that
the gluon dressing $Z(k^2)$  vanishes at the origin, while the ghost $G(k^2)$ is highly singular,
see e.g. Refs.~\cite{AlkoferRunning,IRGluonProp_AlkoferSmekal,SmekalAnnPhys,IRGreen_FewBody2012}.
In the deep IR region, $k\le \epsilon$ the gluon and ghost dressing are predicted to behave as
\begin{equation}
Z(k^2\le \epsilon^2) \sim (k^2)^{2\kappa}; \quad G(k^2\le \epsilon^2) \sim (k^2)^{-\kappa}, \label{limit}
\end{equation}
where $\kappa $ varies in the interval $\kappa\simeq  0.45 - 0.92$ and $\epsilon\sim 0.1-1.0$ MeV/c.
In this  region,  we  force "by hand" the ghost and gluon propagator  to follow Eq.~(\ref{limit}),
 i.e. they do not change during the iteration procedure. In other words, the iteration starts
 at $k > \epsilon$.  We choose $\kappa = 0.45$ and $\epsilon= 0.1$ MeV. Since $\epsilon$ is extremely small,
the  constrains in Eq.~(\ref{limit})   do  not affect the loop integrations
  and they are  not substantial at all in our further calculations.
  Second, the Gaussian form of the interaction  kernels~(\ref{ff1})-(\ref{ff2})  assures a good
 convergence of the iteration procedure. In principle,
 it suffices  to employ  a relatively small  mesh ($  48-64$ Gaussian nodes) to find a stable solution
 of Eqs.~(\ref{eucl1})-(\ref{eucl2}). However, in the subsequent calculations of the propagators
 in complex plane, the tDS equation solution along the real axis $q$
  is used for complex values of $k^2$ for which
 the integrands become highly  oscillating functions at large values of Im~$k^2$.
To assure a good accuracy of numerical  calculations  in this case one needs to have the solution
of the tDS equation along the real axis in a sufficiently dense Gaussian
mesh. For this sake, the whole interval  $q=[0..q_{max}]$ is divided in to three parts:
 (i) $[0\le q\le \epsilon]$ with 16 Gaussian nodes. In this interval the ghost and gluon  dressing functions
   are   taken in accordance with  Eqs.~(\ref{limit}),
 (ii) $[ \epsilon \le q \le 1.1{\rm~GeV/c}]$ is the  interval around the maximum of the gluon propagator.
 Here the Gaussian mesh is taken to consist on 156 nodes,
(iii) in the remaining interval  $[1.1{\rm~GeV/c}\le q \le  q_{max}]$   the Gaussian mesh with 120-156 nodes
is used. The maximum value $q_{max}$ is chosen so that the integrands~(\ref{eucl1})-(\ref{eucl2}) are independent
of $q_{max}$. In our case, the value $q_{max}=5$~GeV/c is sufficient to assure a good accuracy of the solution.
By iterating Eqs.~(\ref{eucl1})-(\ref{eucl2}), we fit the parameters $D_i$ and $\omega_i$ of the
kernels~(\ref{ff1})-(\ref{ff2}) to obtain a reliable agreement with the
lattice QCD results~~\cite{GhostLatticeMishaPRD,BornyakovLattice}.  The renormalization constants
$\tilde Z_3$ and $Z_3$ are defined at the renormalization point $\mu=2.56$ GeV and $\mu=3.0$ GeV
respectively. With the set of parameters
$D_1=1.128\ {\rm GeV^2/c^2}$, $D_2=0.314\ {\rm GeV^2/c^2}$, $D_3=95\ {\rm GeV^2/c^2}$,
$\omega_1=0.7\ {\rm GeV/c}$, $\omega_2=2.16\ {\rm GeV/c}$ and $\omega_3=0.55\ {\rm GeV/c}$
the renormalization constants are found to be $\tilde Z_3=1.065$ and $Z_3=1.05$. The corresponding
solution for the ghost and gluon propagators are presented in Fig.~\ref{ghost}.
 It is seen that both, ghost and gluon dressings are smooth, positively defined
 functions not containing any singularity, except for the ghost dressing,
 which according to Eq.~(\ref{limit}),  is singular at the origin. One can conclude that, with the
 chosen set of parameters, the solution of the tDS equation satisfactorily describes the lattice data. This encourages
 us to use the tDS equation along the real $q$ to find solutions  for complex $k$, treating it  as external parameter
 in the tDS equation.
\section{Solutions of the \lowercase{t}DS equation in complex plane}
\label{sec3}
The solution of the tDS equation along the positive real axis of momenta
is generalized to complex values of $k^2$, needed  to solve the tBS equation for bound states.
The tBS equation is defined in a restricted complex domain of Euclidean space,
  which is determined by the propagators of the constituents. Usually this
  momentum region is displayed as the dependence of the imaginary part of the constituent gluon
   momentum squared, Im\,$k^2$,  on
  its real part, Re\,$k^2$, determined by  the tBS equation.   In terms of the relative  momentum
  $k_{rel}$ of the  two dressed gluons residing in a glueball, the corresponding dependence
  is
  \begin{eqnarray}
  k^2 = -\displaystyle\frac{M_{gg}^2}{4} + k_{rel}^2 \pm i M_{gg} k_{rel} \label{parab}
  \end{eqnarray}
   determining   in the Euclidean complex momentum plane
   a parabola~ Im\,$k^2=\pm~M_{gg}\sqrt{{\rm Re}\,k^2~+~\frac{M_{gg}^2}{4}}$ with  vertex at
  Im\,$k^2=0$ at Re\,$k^2 =-M_{gg}^2/4 $  depending  on the
  glueball mass $M_{gg}$. The symmetry axis is the Re\,$k^2$
  axis, i.e. the parabola extends to Re\,$k^2\to\infty$.

 It should be noted that, for the quark-antiquark bound states
 the use of the complex rainbow solution  in to the tBS equation
provides  an amazingly  good description of  many properties of light mesons
 (masses, widths, decay rates etc.,
 cf.~\cite{Maris:2003vk,Holl:2004fr,Jarecke:2002xd,Krassnigg:2004if,Roberts:1994hh,Maris:2000sk,Maris:1999bh,ourFB}).
 However,   for heavier
 mesons  the quark propagators possess pole-like
 singularities~\cite{OurAnalytical,dorkinBSmesons} which hamper the numerical procedure in
 solving  the tBS equation. An analogous situation can appear for the complex solution
 of the gluon and ghost propagators. Hence, a more detailed investigation of the
  gluon dressing functions in the complex Euclidean plane is required. There are some
  considerations, based on studies of the gauge fixing problem, according to which
   the gluon propagator contains complex conjugate poles in the negative
 half-plane of squared complex momenta $k^2$~\cite{ZwangerANalit,Stingl,CucchieriAnalit}.
 The knowledge of the nature of singularities and their exact location
 in the complex plane is of a great importance since it will allow one
 to develop   effective  algorithms adequate for
 numerical calculations. For instance, if one determines exactly the
 domain of analyticity of the propagator functions, one can take advantage of the fact that
 any  analytical function can always be approximated by rational
 complex  functions \cite{walsh}; then, one can   parametrize the
 integrand  in the tBS equation by simple functions which allow ones to carry out some
 integrations analytically.

There are several possible procedures (cf. Refs.~\cite{analiticalFischer,GluonAnalyticalFisherPRL2012})
 of how to obtain a complex solution of the tDS equations once the equation has
been solved for real and spacelike Euclidean momenta.
 First, one can use the so-called shell method. This method acknowledges the fact that for fixed external
momentum $k^2$ the relative momentum $(p-q)^2$ samples only a parabolic domain in the complex momentum plane.
Therefore, one starts with a sample of external momenta on the boundary of a typical domain very close to
the real positive momentum axis. The tDS equations are then solved on this boundary, while the
interior points are obtained by interpolation. In the
next step, a slightly larger parabolic domain is used, with points in the interior given by the previous solution.
This way one extends the solution of the tDS equations step by step further away from the Euclidean
result into the whole complex plane. A shortcoming of the method is that there is an accumulation
 of numerical errors at each step of the calculations.

A second option  is to deform the loop integration path itself away from the real positive
$k^2$ axis~\cite{MarisComplexDSE,OurAnalytical}.  This can be done by deforming the integration
contour and solving the integral equation along this new contour.
For complex momenta $k$, one has to solve the integral equation along a
deformed contour in the complex plane.
In practice, one changes the integration contour by rotating it in the complex plane,
multiplying both the internal and the external variable by a phase factor $e^{i\phi}$, so one gets the
complex variables $k=|k| e^{i\phi}$  and  $q=|q| e^{i\phi}$ and solves the tDS equation along the rays $\phi=const$.
This method works quite well in the first quadrant, $\phi \le \pi/2$,
but fails at $\phi > \pi/2$, see e.g. Ref.~\cite{OurAnalytical,dorkinBSmesons}.
 This is because along the rays $\phi=const$
all the values of $|k|$, from $|k|=0$  to $|k|\to\infty$ contribute to the tDS equation,
  even if  one needs the solution only in a  restricted area of the parabola
 Im\,$k^2 < 0$. Consequently, numerical instabilities are inevitable at $\phi > \pi/2$.

The third method, which we use in this work, consists in  finding a solution
to the integral equations in a straightforward way from the tDS equation along the real $q$
  on a complex grid for the external momentum $k$ inside and in the  neighbourhood
of  the parabola (\ref{parab}).
  As in the previous case, numerical instabilities  can be caused by oscillations of
  the exponent ${\rm e}^{-(k-q)^2/\omega^2}$ and of the Bessel functions $I_{1,2}^{(s)}(2kq/\omega^2)$
  at large $|k^2|$.  However, one can get rid of such a numerical problem by taking into account
  that parabola (\ref{parab}) restricts only a small portion of the complex plane at Re\,$k^2<0$, where
  the numerical problems are minimized.
 For positive values of Re\,$k^2>0$, where $|k^2|$ can be large,  the tBS wave function
 of a glueball  is expected to decrease rapidly with increasing  argument $ k_{rel}$, and
  at $k_{rel}^{max}\sim 3-4 \ {\rm GeV/c}$  to become  negligibly small. So,  one can
 solve the complex  tDS equation at not too large $|k^2|$, where  a reliable calculation
 of the loop integrals in Eqs.~(\ref{eucl1})-(\ref{eucl2}) is still possible. Then one takes advantage
of the fact that, at larger values of
 $k_{rel}$, the highly oscillating integrals in (\ref{eucl1})-(\ref{eucl2}) are
 negligible small or  even vanish at $k_{rel}\to\infty$, in accordance with the Riemann-Lebesgue lemma.
Consequently, such integrals can be safely neglected. This  means that it suffices to investigate the behaviour
of the gluon and ghost propagators for a parabola with $k_{rel}^{max}\sim 3-4\ {\rm GeV/c}$.
In what follows we analyze the analytical properties of
the dressing functions $Z(k)$ and $G(k)$ in the complex domain of parabola
 from $Re k^2$ corresponding to glueball masses
$M_{gg} \le 5\, {\rm GeV/c^2}$ up to Re\,$k^2$ corresponding to $k_{rel}^{max}\sim 3.5\ {\rm GeV/c}$.

\subsection{Analytical structure  of the gluon and ghost propagators in complex Euclidean space}
\label{analyt}
 To determine the analytical properties of $Z(k)$ and $G(k)$ we use a combined
 method~\cite{dorkinBSmesons,OurAnalytical} based on  calculations of  the Cauchy and Rushe integrals.
In a closed domain, the Cauchy integral of an analytical function $f(z)$ vanishes.
 Contrarily, the non-zero Cauchy integral undoubtedly indicates that $f(z)$ is
 singular inside the domain. In this case, to locate and investigate the nature of the
 singularities one computes the Cauchy integral of  the inverse function $g(z)=1/f(z)$.
 The vanishing Cauchy integral of the inverse function  means that
 $g(z)$ is analytical in the considered domain.
 Consequently, one concludes that the singularities of $ f(z)$ can be solely
 of the pole-type.  Evidently, the positions
 of such poles coincide with the positions of the zeros of $g$. The zeros of $g$
 can be found by the Rushe theorem~\footnote{Rouch\' e integral of an analytical
  complex function $g(z)$ on a closed contour $\gamma$  is defined by
 $\frac{1}{2\pi i}\oint\limits_\gamma\frac{g'(z)}{g(z)}d z$.}, according to which
 the  Rouch\' e integral must be an integer,   exactly equal
 to the number of zeros inside the domain.
 Our  calculations show that both,  Cauchy integral of $G(k^2)$ and Cauchy integral of $Z(k^2)$,
 are different from zero, i.e. $G(k^2)$ and $Z(k^2)$ are  singular inside the parabola. Then,
 the further strategy of finding  these singularities  is as follows:\\
(i) Consider consecutively the dressing functions $G(k^2)$ and  $Z(k^2)$.
  Choose  a contour inside the parabola and compute  the Cauchy  integral
 of $G(k^2)$ (or $Z(k^2)$). If the integral is zero, we choose another contour nearby the previous one
 and repeat the calculations until a non-zero integral is encountered.
 Check whether the singularities here are of pole-type, i.e. compute  the Cauchy  integral
 of the inverse, $G(k^2)^{-1}$ (or $Z(k^2)^{-1}$), which must be zero if singularities are
 isolated poles.\\
(ii) Compute the Rouch\'e   integral of    $G(k^2)^{-1}$ (or $Z(k^2)^{-1}$).
Since the inverse function has been found to be analytical,
such an integral, according to the Rouch\'e's theorem,  gives exactly the number
of  zeros  inside the  contour.\\
(iii) Squeeze the contour and repeat items (i)-(ii), keeping the zero inside, until an
isolated  zero of  $G(k^2)^{-1}$ (or $Z(k^2)^{-1}$) is located with a desired accuracy. The corresponding
integrals  around such isolated poles $k_{0i}^2$  read as
\begin{eqnarray}&&
 \frac{1}{2\pi i}\oint\limits_\gamma \left[  G(k^2) \left (Z(k^2)\right)\right]d k^2 =
  \sum_i res\left[G(k_{0i}^2) \left (Z(k_{0i}^2)\right)\right],   \label{ucl}\\ &&
  \frac{1}{2\pi i}\oint\limits_\gamma G^{-1}(k^2) \left(Z^{-1}(k^2)\right) dk^2 =0. \label{zero}\\ &&
 \frac{1}{2\pi i} \oint\limits_\gamma
\frac{  \left[ G^{-1}(k^2) \left(Z^{-1}(k^2)\right) \right]'_{k^2} }{G^{-1}(k^2)
 \left(Z^{-1}(k^2)\right)}dk^2 =N_{G(Z)},
 \label{resids}
 \end{eqnarray}
 where $N_{G(Z)}$ is the number of poles  in the domain enclosed by the contour $\gamma$
(an effective algorithm
for numerical evaluations of Cauchy-like integrals can be found, e.g. in Ref.~\cite{ioak}).
   In such a way we find the   poles of $G(k^2)$ and $Z(k^2)$  together with their residues relevant
 for further calculations. In subsequent numerical calculations of integrals, involving functions with pole-like
 singularities, one can use the following theorem: if a complex function $f(z)$ possesses isolated poles,
 then it can be represented in the  form
 \begin{equation}
 f(z) = \widetilde f(z) + \sum_i\frac{{\rm res} [f(z_{0i})]}{z-z_{0i}},
\label{main}
 \end{equation}
 where $\widetilde f(z)$ is analytical  within
 the considered domain and, consequently, can be computed as
 \begin{equation}
 \widetilde f(z)
  =\frac{1}{2\pi i}\oint\limits_{\gamma}\frac{\widetilde f(\xi)}{\xi -z} d\xi
 =  \frac{1}{2\pi i}\oint\limits_{\gamma}\frac{f (\xi)}{\xi -z} d\xi .
  \label{f1}
 \end{equation}

  Note that a good numerical test of the performed calculations is the following procedure.
  Enclose a few poles by a larger contour and
  ensure that  the Cauchy integral of  $G(k^2)$ or $Z(k^2)$ is different from zero and that
 the   Rouch\'e integral of the inverse, $G(k^2)^{-1}$ or $Z(k^2)^{-1}$, is an integer
  equal to the number of enclosed poles.
 Note that the Cauchy integral of $G(k^2)$ or $Z(k^2)$ in this case
 must coincide with   the sum of individual residues of the isolated poles.

\subsection{Pole structure of the dressing functions}
\label{poleStruc}
\begin{table}[!h]
\caption{The pole structure of  the gluon, $Z(k^2)$, and ghost, $G(k^2)$, dressing functions.
 The pole positions $(Re\ k_{0}^2, Im\  k_{0}^2)$ and the corresponding residues are in
units of  ${\rm (GeV/c)}^2$.
Only the first, four self-conjugated poles on $k^2$
close to the parabolas~(\ref{parab}), see also Fig.~\protect\ref{poles}, are presented.}
\begin{tabular}{lccccc} \hline\hline
        Gluons            &      1     &        2     &      3      &  4  \\ \hline
   $k_{0i}^2$           &  (-3.52, $\pm$ 6.97)     & (-1.975, $\pm$ 2.05)&(-0.605,$\pm$ 6.02) &(0.11,$\pm$ 0.61)\\
  res[$Z(k_{0i}^2$ ]    & (-0.0536, $\mp$ 0.01755) & (0.051,$\mp$ 0.081) & (0.79, $\mp$  0.079) & (0.589,$\pm$ 0.0791)
 \\ \hline\hline\\
  Ghosts         &             1                &        2              &            3                                &  4  \\ \hline
  $k_{0i}^2$            & (-2.47, $\pm$ 7.37)   &  (-1.915,$\pm$ 4.15)  &      (-0.687, 0.0)       &--\\
   res[$G(k_{0i}^2$ ]   & (0.4667 $\pm$ 0.036)  & (0.494,  $\pm$ 0.082) &      (0.956,$\mp$ 0.0)   & --\\
 \\ \hline\hline
\end{tabular}
\label{polePosition}
\end{table}

\begin{figure}[h]
\includegraphics[scale=0.5,angle=0]{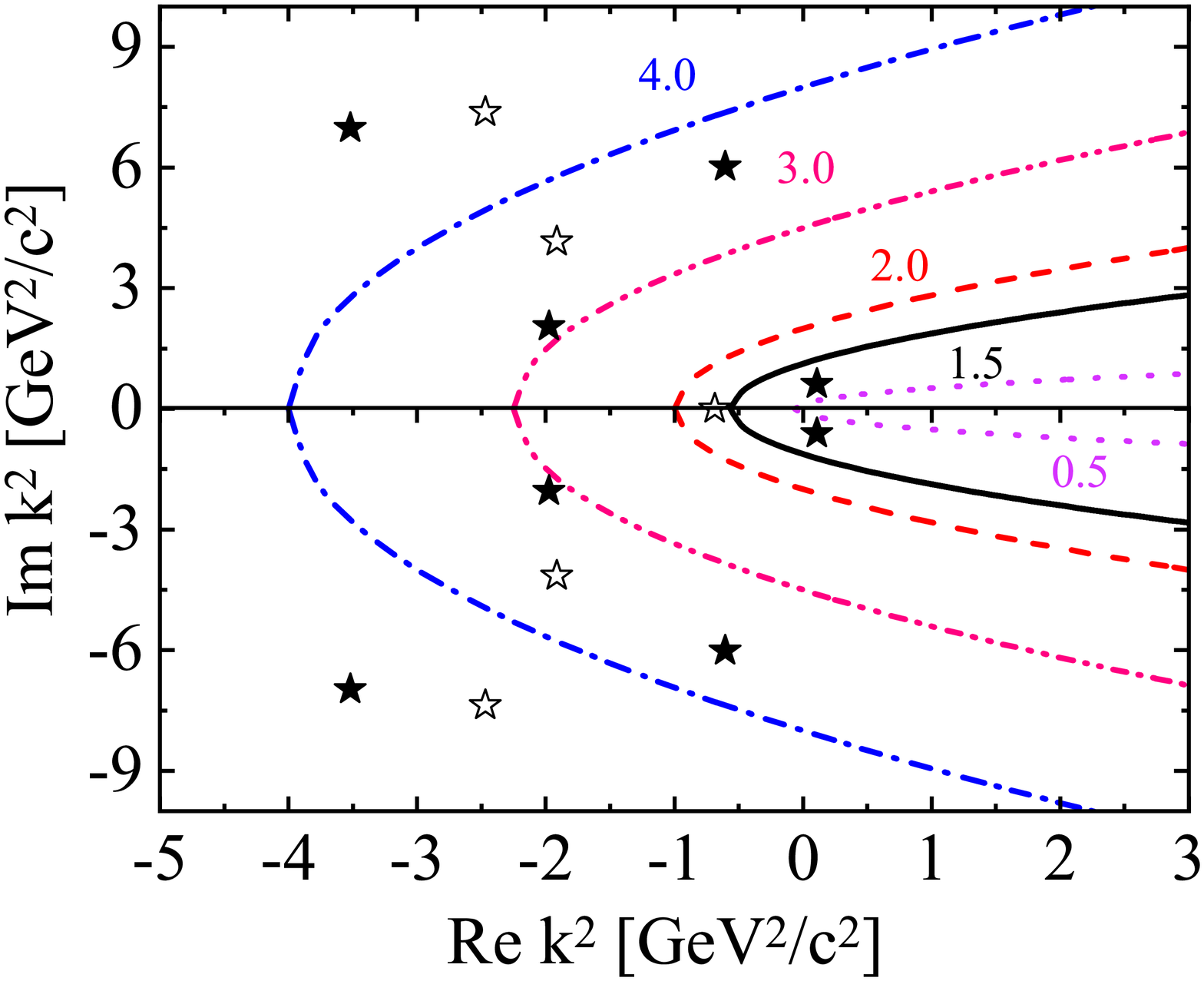}
\caption{(Color online)  Positions of few first poles of the gluon
$Z(k^2)$, full stars, and  ghost, $G(k^2)$, open stars, dressing function in the complex $k^2$ plane,
 labeled in correspondence to  the Table~\ref{polePosition}.  The relevant sections of the parabola
  (\ref{parab})  corresponding to
 the glueball bound-state mass $M_{gg}$
 are presented for $M_{gg}=0.5,\, 1.5,\, 2,\, 3$ and $4$  GeV, from right to left.
The tendency is   that, with increasing  glueball bound-state mass $M_{gg}$, more and more
poles are located in the physical area, relevant to the tBS equation.
}
\label{poles}
\end{figure}
Results of our calculations are presented in Table~\ref{polePosition} and Fig.~\ref{poles}.
It is seen that for
$M_{gg}< 5\,{\rm  GeV/c^2}$ all singularities of the gluon dressing $Z(k^2)$
are pairwise complex conjugated. There are two complex conjugated poles
at Re\,$k^2 > 0$ which means that a glueball bound state contains at least two poles, regardless
the glueball mass $M_{gg}$, except for very low values of $M_{gg} < 0.5 \ {\rm GeV}$,  see Fig.~\ref{poles}.
Contrarily, all singularities of the ghosts are located  in the region $ Re\ k^2 < 0$ with one
  real pole  at Re\,$k^2 \approx -0.69\ {\rm GeV^2/c^2}$.

Results of calculations by Eqs.~(\ref{ucl})-(\ref{resids}) provide all the necessary ingredients for
Eqs.~(\ref{main})-(\ref{f1},) allowing to establish easily reliable algorithms
 for  solving  numerically the tBS equation even in the presence of pole-like singularities.

With these calculations our analysis of the pole structure is completed. Let us recall the
prepositions: (i) The tDS equation is solved in an approach similar to the rainbow approximation
with IR part only. The phenomenological
parameters have been adjusted to lattice QCD results.
(ii) The tDS equation, restricted to the momentum range relevant
for  $gg$ bound states, $M_{gg}< 5\,{\rm   GeV/c^2}$.

\section{Summary}
\label{summary}
 We analyse analytical properties of  solutions of the truncated Dyson-Schwinger equation
 for the ghost and gluon propagators  in the Euclidean
 complex momentum domain which is determined by the truncated Bethe-Salpeter equation
 for  two-gluon bound states. Our approach is based on an approximation, similar to the
 rainbow approach for quarks, with effective parameters adjusted in accordance with the
 available lattice QCD data. It is found that, within the suggested approach with only the infrared terms
 in the combined effective vertex-gluon dressing and vertex-ghost dressing kernels,
 the solutions $Z(k^2)$  and $G(k^2)$ are singular in  the whole
 considered domain for  $M_{gg}< 5\,{\rm   GeV/c^2}$, with singularities as
 isolated pairwise complex conjugated poles.
 The exact position of the poles and the corresponding residues of the propagators
 can be found by applying Rouch\'e  theorem and computing the Cauchy integrals.

 The position of the few first poles and the corresponding residues are found with  good accuracy
  to be used in further calculations based on the Bethe-Salpeter  equation. It is also found that, with
 only the effective infrared  term in the parametrization of the combined  vertex-gluon
and vertex-ghost kernels, the ghost dressing function exhibits a pole on the negative real axis.
 The performed analysis is aimed at elaborating  adequate
 numerical algorithms to solve the truncated Bethe-Salpeter equation in presence of singularities and to investigate the properties of
glueballs, e.g. scalar and pseudoscalar glueball states.

\section*{Acknowledgments}
This work has been  supported  by the National Natural Science Foundation of
China (grant No. 11575254) and Chinese Academy
 of Sciences President's International Fellowship Initiative
 (No. 2018VMA0029).
  LPK appreciates the warm hospitality at the
 Institute of Modern Physics, Lanzhou, China.
 The authors gratefully acknowledge the fruitful discussions with N. I. Kochelev and
 S.M. Dorkin.

 \appendix
 \section{}
\label{app}
\begin{figure}[hbt]
\includegraphics[scale=0.5,angle=0]{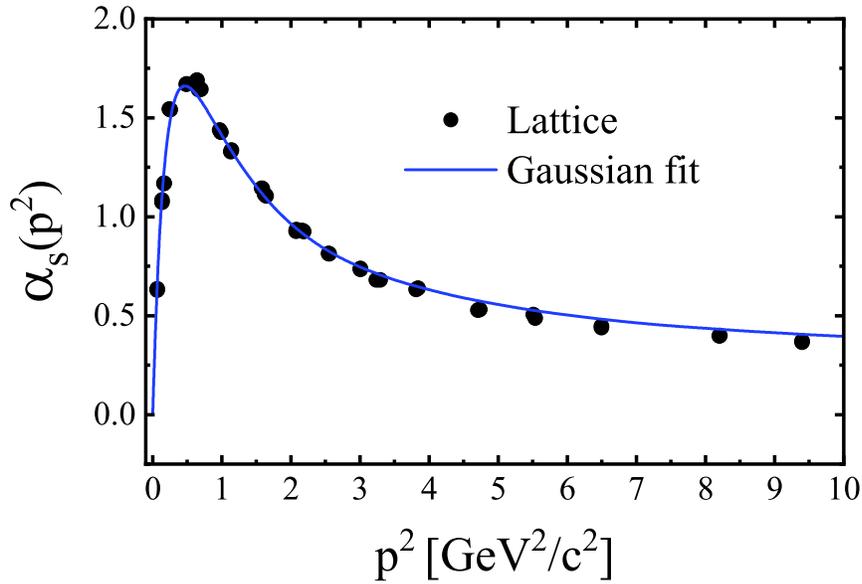}
\caption{(Color online) The fit of the  nonperturbative running coupling  $\alpha_s(p)$
by Eq.~(\ref{parametr}) (solid line) vs. the results of lattice QCD calculations~\cite{GhostLatticeMishaPRD} (filled circles).}
\label{running}
\end{figure}

 The   model interaction kernels in the rainbow approximations
 is   inspired essentially  by the behaviour of the running
 coupling~(\ref{alphas}) in the IR region, which now is available from the lattice QCD
 data~\cite{GhostLatticeMishaPRD}. In order to facilitate the  calculations,
 the explicit expressions  for the kernels are taken in form of Gaussian terms. Accordingly,
it is preferably to have  the parametrization of the running coupling  also in such a form.
 Usually, in original publications of the lattice QCD results one employs parametrizations
 to fit   data as a sum of several multipole  terms, cf.~\cite{GhostLatticeMishaPRD,BornyakovLattice}.
For our purpose we have to refit the data within another, Gaussian-like formula.
  Here below we present a fit   for the
  running coupling~(\ref{alphas}) in form of a sum of several Gaussian terms with
  fitting parameters found from a Levenberg minimization procedure.
  Such a parametrization serves as a guideline  in
  choosing the form of the effective kernels~(\ref{ff1})-(\ref{ff2}).

 \begin{equation}
 \alpha_s(p^2) = p^2 \sum_{i=1}^5 A_i  {\rm e}^{-a_i p^2}. \label{parametr}
 \end{equation}
\noindent
 The minimization procedure converged to a set of parameters listed  in Table~\ref{tabAlpha}, which
 provide a fit of lattice QCD data  presented in Fig.~\ref{running}.

  \begin{table}[!ht]
  \caption
{The parameters
 $A_i$ and  $a_i$ (in $[ {\rm \left( GeV/c\right )^{-2}}]$) for the effective parametrizations, Eq.~(\ref{parametr}), of the
 lattice QCD results~\cite{GhostLatticeMishaPRD}.}
\begin{tabular}{l| c l lcc c c ccc } \hline\hline  % after \\: \hline or \cline{col1-col2} \cline{col3-col4} ...
     &\phantom{p}   & 1 &\phantom{p} & 2&\phantom{p}& 3&\phantom{p}&4 &\phantom{p}&5\\ \hline
    $A_i  $     &\phantom{p}   & 4.546 &\phantom{p} & 0.840&\phantom{p}& 0.146&\phantom{p}&2.472 &\phantom{p} &6.87 \\ \hline
    $a_i$     &\phantom{p}   & 1.804 &\phantom{p} & 0.636&\phantom{p}& 0.196 &\phantom{p}&3.45&\phantom{p} &4.51 \\\hline
                     \hline
\end{tabular}
\label{tabAlpha}
 \end{table}

\end{document}